\documentclass {camera} 
\usepackage {epsfig} 
\usepackage {color} 
\usepackage {amssymb} 
 
\renewcommand {\section}[1]{} 
\renewcommand {\subsection}[1]{} 
\def\ii{{\rm i}}
 \renewcommand{\vec}[1] {{\mathbf{#1}}}
 \newcommand{\vt} {\vartheta}
 \newcommand{\bea} {\begin{eqnarray}}
 \newcommand{\eea} {\end{eqnarray}}
 \newcommand{\beann} {\begin{eqnarray*}}
 \newcommand{\eeann} {\end{eqnarray*}}
 \newcommand{\labs} {\left\vert}
 \newcommand{\rabs} {\right\vert}

 \newcommand{\lrb} {\left(}
 \newcommand{\rrb} {\right)}
 \newcommand{\lcb} {\left\{}
 \newcommand{\rcb} {\right\}}
 
 \newcommand{\rab} {\right\rangle}

\begin{document}
\begin{picture}(1,1)
\put(-10,20){
\begin{minipage}{10cm}
{\small
``Nanotubes and Nanostructures''	\\
S. Bellucci (Ed.)			\\
SIF, Bologna, 2001			\\	
ISBN: 88-7794-291-6			\\
}
\end{minipage}
}
\end{picture}

\vspace*{.1cm}

\title {A molecular wire sandwiched between nanotube leads: analytic results
for the conductance}
\author {G. Cuniberti, G. Fagas, \And K. Richter}
\organization {Max-Planck-Institut f{\"u}r Physik komplexer Systeme, 
\\ N{\"o}thnitzer Stra{\ss}e 38, D--01187 Dresden, Germany
}
\maketitle
\begin{abstract}
 Analytic results for the conductance of a molecular wire attached to mesoscopic tubule leads are obtained. They permit to study linear transport 
 in presence of low dimensional leads in the whole range of parameters. 
 In particular contact effects can be addressed in detail.
By focusing on the specificity of the lead--wire contact, we show that the geometry of this hybrid system supports a mechanism of channel selection, which is a distinctive hallmark of the mesoscopic nature of the electrodes.
\end{abstract}

\section{Introduction}

The nanometer length domain, the fate of the transistor integration length, is unavoidably rising a new variety of phenomena concerning the r{\^o}le of quantum effects on electronics at the molecular scale~\cite{Taur97}. 
Although molecular materials for electronics are already realized~\cite{CDLFBLSKL00,SKBB00,Ziemelis98,Tour96}, real molecular scale electronic devices \cite{EL00,TKS98,G-GMLOL97} still have to cope with challenges in utilization, synthesis, and assembly~\cite{Landauer96c}.
The idea of molecular rectification proposed since 1974~\cite{AR74a} 
found experimental evidence only 20 years later~\cite{WB93,MSA93}.
Concerning theory, conventional methods employed for characterizing transport properties in microelectronic devices, such as the Boltzmann equation~\cite{Fischietti84}, can no longer be applied at the molecular scale.
Here transport properties have to be calculated by using full quantum mechanical approaches.

Electron transmission through molecular and supramolecular interfaces was already the object of intensive theoretical investigation in the last decade due to the electron transfer phenomena underlying the use of scanning tunneling microscopes.
More recently, studies of transmission properties of a molecular junction 
contacted to metallic leads 
\cite{RZMBT97,PBdeVD00}
have intensified the interest in the basic mechanisms of conduction across molecular bridges.
In a parallel development the use of carbon nanotube (CNT) networks has been the focus of intense experimental and theoretical activity as another promising direction for building blocks of molecular circuits~\cite{KKKCSRO01,RKJTCL00}.
Carbon nanotubes are known to exhibit a wealth of properties depending on their nano--scale diameter, orientation of graphene roll up, and whether they consist of a single cylindrical surface~(single--wall) or many~(multi--wall)~\cite{SDD98,McEuen00}.
However if carbon nanotubes are attached to other materials, the characterization of contacts~\cite{ADX00,dePGWADR99}
becomes a fundamental issue with respect to the employment of tubes as elements of molecular circuits. This could be the case when a carbon nanotube is attached to a single molecule or a molecular cluster with a privileged direction along the current flow, namely a molecular wire.

The archetype of a molecular device can be viewed as a donor and acceptor lead coupled by a molecular wire bridge. 
In such systems the traditional picture of electron transfer between donor and acceptor species is re--read in terms of a novel view in which a molecule can bear an electric current~\cite{Nitzan01}.
In the usual treatment of transport through molecular wires, the attached leads are represented by a continuum of free or quasi--free states, mimicking the presence of large reservoirs provided by bulky electrodes. 
However such an assumption may become inadequate when considering
leads with lateral dimensions of the order of the bridged molecule, such as
carbon nanotubes.
The latter have been recently used for enhancing the resolution of scanning probe tips~\cite{WJWCL98}. In a similar configuration the fine structure of a twinned DNA molecule has been observed~\cite{NKANHYT00}. 
One can argue that the presence of such {\it mesoscopic leads} strongly influences the conductance properties across the molecular bridge.

\begin{figure}[t]
\phantom{.}\vspace*{0.1cm}
 \centerline{\epsfig{file=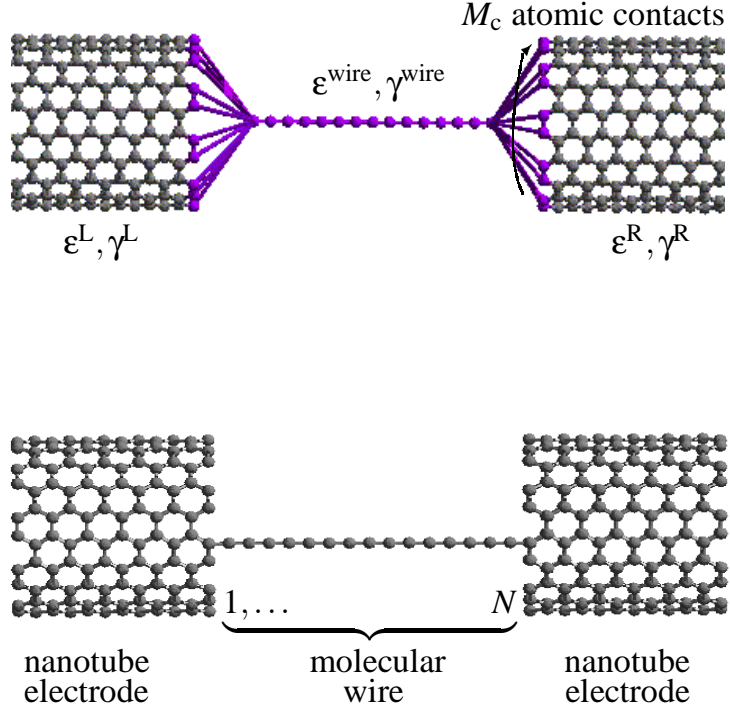, width=.60\linewidth}}
\phantom{.}\vspace*{1cm}
\caption{\label{fig:sketch} Scheme of the molecular wire-carbon nanotube hybrid with single (bottom) and multiple (top) contacts. In this paper, on-site energies $\epsilon^{\alpha={\rm L,R,wire}}$ are fixed to zero. }
 \end{figure}

This paper addresses the influence of the molecular wire--electrode contact on the conductance when the structure of electrodes plays an important r{\^o}le.
In order to isolate contact effects, we treat analytically a minimal model where the tubular leads are described by a rolled square lattice and the molecule by a homogeneous linear chain. 
This system supports distinct transport properties depending on the number and strength of contacts between the molecular bridge and the interface as well as on the symmetry of the channel wave function transverse to the transport direction. 
The validity of these analytical findings are then tested for real tubular systems by treating CNT leads (figure~\ref{fig:sketch}.) via a numerical recursive Green's function approach.
Such a study delivers additional insight to recent numerical results on carbon nanotubes/molecular wire hybrids~\cite{FCR01a}.
In particular, we demonstrate that the configuration when only one molecule--lead contact is activated gives rise to complex conductance spectra exhibiting quantum features of both the molecule and the electrodes; on the other hand multiple contacts provide a mechanism for transport channel selection leading to a scaling law for the conductance and allowing for its control. Channel selection also highlights the r{\^o}le of molecular resonant states by suppressing details assigned to the electrodes.

\section{Method}
The hamiltonian of the full system $H = H_\mathrm{tubes} + H_{\rm wire} + H_{\rm coupling}$
reads
\begin{eqnarray}
\label{eq:hamiltonian}
 H &=& 
 \sum_{\alpha={\rm L,R,wire}} \; 
 \sum_{n^{\phantom{\prime}}_{\alpha},n_{\alpha}^\prime} 
 t^\alpha_{n^{\phantom{\prime}}_{\alpha},n_{\alpha}^\prime}
 {a_{n_\alpha^{\phantom{\prime}}}^{\alpha \dagger}}
 a_{n_\alpha^\prime}^{{\alpha \phantom{ \dagger}}}
 \\ && 
 - \sum_{ m_{\rm L} \le M_{\rm c}} \Gamma_{m_{\rm L}}
 \lrb
 {{a_{m_{\rm L}}^{{\rm L} \dagger}}} a_1^{{\rm wire \phantom{ \dagger}}}
 + {\rm h.c.} \rrb
 \nonumber
 - \sum_{m_{\rm R} \le M_{\rm c}} \Gamma_{m_{\rm R}} 
 \lrb
 {a_{m_{\rm R}}^{{\rm R} \dagger}} a_N^{{\rm wire \phantom{ \dagger}}}
 + {\rm h.c.} \rrb
 \nonumber
\end{eqnarray}
where the matrix element $t^\alpha_{n^{\phantom{\prime}}_{\alpha},n_{\alpha}^\prime} = 
\varepsilon^{\alpha}_{n_{{\alpha}}} \delta_{n^{\phantom{\prime}}_{{\alpha}},n_{{\alpha}}^\prime} - \gamma^\alpha_{\left\langle n^{\phantom{\prime}}_{\alpha},n_{\alpha}^\prime \right\rangle}$ contains the on--site energy of each of the $n_ {\rm wire}=1,\dots, N$ chain--atoms, $\varepsilon^{\rm wire}$, the orbital energy relative to that of the lead atoms, $\varepsilon^{\rm L,R}$, and $\gamma^{\rm L,R}$, $\gamma^{\rm wire}$, and $\Gamma$ are nearest neighbour hopping terms between atoms of the left or right leads, molecular bridge, and the bridge/lead interface, respectively.
Note that $n_{\rm L,R}$ is a two--dimensional coordinate spanning the tube lattice.
Summations over $m_{\rm L}$ and $m_{\rm R}$ run over interfacial end--atoms of the leads.
In general, 
there are $M$ such atomic positions, defining the perimeter of the tube ends.
the number of hybridization contacts range between $M_{\rm c}=1$ (SC) and $M_{\rm c}=M$ (MC).
The square lattice tubes (SLT)
are obtained by imposing periodic boundary conditions on the longitudinal
cuts parallel to the lattice bonds of length $a$. 
In the case of CNT, when the graphene honeycomb lattice is rolled along the
lattice bonds, armchair single wall $(\ell,\ell)$ nanotube are obtained. In this case
$M=2\ell$.

In order to derive transport properties, we make use of the Landauer theory~\cite{IL99} which relates the conductance of the system to an independent--electron scattering problem~\cite{FG99}.
The electron wavefunction is assumed to extend coherently across the device and the two--terminal, linear--response conductance at zero temperature, $g$, is simply proportional to the total transmittance for injected electrons $T(E_{\rm F})$ at the Fermi energy $E_{\rm F}$:
\begin{equation}
\label{eq:linear-zero-temperature-conductance}
g = \frac {2 e^2} h T(E_{\rm F}).
\end{equation}
The factor two accounts for spin degeneracy. The transmission function can be calculated from the knowledge of the molecular energy levels, the nature and the geometry of the contacts. 
It is given by 
\bea
T(E)=
\sum_{j_{\rm L},j_{\rm R}} \labs S_{j_{\rm L} j_{\rm R}} \rabs^2= 
{\rm Tr} \lcb S S^\dagger \rcb, 
\label{eq:transmission-function}
\eea
where $j_{\rm L},j_{\rm R}$ are quantum numbers labeling open channels for transport which belong to mutually exclusive leads, in our case the two semi--infinite perfect nanotubes. The molecular system attached acts as a scatterer, and $S$ is the corresponding quantum--mechanical scattering matrix. The quantity $\labs S_{j_{\rm L} j_{\rm R}}\rabs^2$ is the probability that a carrier coming from, say, left of the scatterer in the transversal mode $j_{\rm L}$ will be transmitted to the right in the transversal mode $j_{\rm R}$. The sum in~(\ref{eq:transmission-function}) is restricted to transversal modes whose energy is smaller than $E_{\rm F}$.

\subsection{Calculus of the transmission after Bardeen}
In the linear regime one can write the zero--temperature transmission as
\bea
\label{eq:Bardeen}
T(E) = 
\sum_{i_{\rm L},i_{\rm R}} 
\labs {\cal T}_{i_{\rm L},i_{\rm R}} \rabs^2 
\delta \lrb E-E_{i_{\rm L}} \rrb
\delta \lrb E-E_{i_{\rm R}} \rrb ,
\eea
by considering the zeroth order states confined to the left and the right electrodes $\labs i_{\rm L} \rab$ and $\labs i_{\rm R} \rab$, respectively.
${\cal T}$ is the corresponding transition operator whose form depends on the
details of the confinement.
Note that these confined states are {\it not} the scattering states labelled by $j_{\rm L}$ and $j_{\rm R}$ in equation~(\ref{eq:transmission-function}) which are the eigenstates of the exact hamiltonian.
Equation~(\ref{eq:Bardeen}) is basically derived within a transfer hamiltonian
treatment {\it {\`a}
la} Bardeen (weak coupling assumption)~\cite{Bardeen61}.
It has been initially developed by Caroli {\it et al.}~\cite{CCNS-J71}, and
more recently applied to molecular wires by Mujica {\it et
al.}~\cite{MKR94a,MKR94b}. 
The relationship between the scattering matrix and the transfer hamiltonian approaches has been extensively worked out in molecular systems~\cite{Nitzan01,HRHS00} showing {\it de facto} their equivalence.
This fortunate fact let us to make use of
the formal result of the Bardeen treatment in the broader applicability context of the Landauer approach (which holds also in the regime of perfect transmission).
In fact the transmission function introduced in equation~(\ref{eq:transmission-function}) is exactly the same as used in equation~(\ref{eq:Bardeen}).
One can see this by writing down the Green's function matrix of the
problem 
$\vec{G}^{-1} = \vec{G^{\rm wire}}^{-1} + \vec{\Sigma_{\rm L} + \Sigma_{\rm R}}$
written in terms of the bare wire Green's function and the self-energy
correction due to the presence of the leads. Making use of the Fisher--Lee
relation~\cite{FL81} one can finally write 
\bea
\label{eq:transm-mit-spectral-densities}
T = {\rm Tr} \lcb \vec{\Delta}^{\rm L} (E) {\vec{G}} (E) \vec{\Delta}^{\rm R} (E) \vec{G}^\dagger (E) \rcb = 4 \Delta^{\rm L}_{11} (E) \Delta^{\rm R}_{NN} (E) \labs G_{1N} \lrb E \rrb \rabs^2 ,
\eea
where $\vec{\Delta}^{\alpha} (E) = \ii {\lrb {\vec{\Sigma}}^\alpha -{{\vec{\Sigma}}^\alpha}^\dagger \rrb } (E + \ii 0^+)$ contain only one non--zero element ($\Delta^{\rm L}_{11}$ and $\Delta^{\rm R}_{NN}$ for the left and right lead, respectively), due to the geometry. 
The rhs of equation~(\ref{eq:transm-mit-spectral-densities}) coincides with the formula provided within transfer hamiltonian schemes~\cite{MKR94a,MKR94b}. 
The matrix element $\Delta^{L (R)}$ is the left (right) lead spectral density which is related to the semi--infinite lead Green's function matrix ${G}_{\rm lead}$.
It is minus the imaginary part of the lead self--energy (per spin)
\bea
\label{eq:spectr-density}
\Sigma^{\alpha=L,R}
= \sum_{m_{\alpha}, m_{\alpha}^\prime} 
 \Gamma^{\phantom{*}}_{m^{\phantom{\prime}}_{\alpha}} 
 \Gamma^*_{m^\prime_{\alpha}} 
 {G}_{\rm lead} \lrb m_{\alpha} , m_{\alpha}^\prime \rrb
 .
\eea
Finally one has to calculate from the Green's function
\bea
\label{eq:bare-molel-green-matrix}
\lrb {G^{\rm wire}}^{-1} \rrb_{n n^{{\prime}} } 
= \lrb E + \ii 0^+ - H_{\rm wire} \rrb_{n n^\prime}
= \lrb E + \ii 0^+ - \varepsilon^{\rm wire }_n\rrb\delta_{n n^{{\prime}} } - 
\gamma^{\rm wire}_{\left\langle n, n^\prime \right\rangle } ,
\eea
the Green's function matrix element $G_{1N}$ in
equation~(\ref{eq:transm-mit-spectral-densities}).
It is given as a matrix element referring to the two $N$--atom--molecule ends and is, computationally, an $N \times N$ matrix inversion.
Since only the molecular--end on--site energies are perturbed by the interaction with the leads via the self--energy $\Sigma^{\alpha}$, some general conclusions can be drawn without the need of an explicit computation of $G_{1N}$.

The Green's function $G_{1N}$, in equation~(\ref{eq:transm-mit-spectral-densities}), reads
\bea
\label{eq:dressed-G1N}
G_{1 N} = \frac {G^{\rm wire}_{1 N}}{
\lrb 1 - \Sigma^{\rm L}_{11} G^{\rm wire}_{1 1} \rrb
\lrb 1 - \Sigma^{\rm R}_{NN} G^{\rm wire}_{N N} \rrb
- 
\Sigma^{\rm L}_{11} \Sigma^{\rm R}_{NN}
\lrb G^{\rm wire}_{1 N} \rrb^2
}.
\eea
The interaction with the leads dresses via the self--energy
$\Sigma^{\alpha}$ the bare molecular wire Green's function
element $G^{\rm wire}_{1 N}$. The latter 
can be calculated analytically in the case of an homogeneous wire 
($\varepsilon^{\rm wire}_n=\varepsilon^{\rm wire}$, 
$\gamma^{\rm wire}_{\left\langle n^{\phantom{\prime}}_{\alpha},n_{\alpha}^\prime \right\rangle}=\gamma^{\rm wire}$).
In fact projecting on the $N$ dimensional molecular wire basis,
the determinant of the bare molecular Green's matrix~(\ref{eq:bare-molel-green-matrix})
factorizes a dimensionless function of only the number of chain atoms, and of the ratio ${\cal E} = (E-\varepsilon^{\rm wire} )/(2 \gamma^{\rm wire})$. 
This leads to a closed form for the molecular contribution in the conductance. Namely,
one can easily check that 
$G^{\rm wire}_{1N}=
{\gamma^{\rm wire}}^{N-1} {\rm det} \lrb G^{\rm wire} \rrb =
{\gamma^{\rm wire}}^{-1} \xi(0)/\xi(N)$,
and
$G^{\rm wire}_{11} =G^{\rm wire}_{NN} ={\gamma^{\rm wire}}^{-1} \xi(N-1) / \xi(N)$,
where the exact form of $\xi$ reads:
\beann
\xi(N) =
{\lrb {\cal E} + \sqrt{{\cal E}^2 -1} \rrb^{N + 1} - \lrb {\cal E} - \sqrt{{\cal E}^2 -1} \rrb^{N + 1}}.
\eeann
After some algebra one finds that $\xi$ possesses the following
recursive property,
\bea
\label{eq:sumrule}
\xi(N) = \frac { \xi^2(N-1) -\xi^2 (0) } {\xi(N-2)}
\eea
which leads us to re--write equation~(\ref{eq:dressed-G1N}) as
\bea
\label{eq:G1Nquasiinbellaforma}
\frac{\xi(0)}{{\gamma^{\rm wire}} G_{1N}} 
= 
{\xi(N)} 
- \lrb \frac{\Sigma_{11}^{\rm L}}{{\gamma^{\rm wire}}} + 
 \frac{\Sigma_{NN}^{\rm R}}{{\gamma^{\rm
 wire}}} \rrb
{\xi(N-1)} 
 + \frac{\Sigma_{11}^{\rm L} \Sigma_N^{\rm R} }{{\gamma^{\rm wire}}^2} 
{\xi(N-2)} .
\eea
This means that the inverse of the Green's function matrix element connecting left and right leads 
can be written as a sum of the terms, representing the inverse of the bare
Green's function matrix elements for a wire of $N$, $N-1$, and $N-2$ atoms.
In the limit of weak contact coupling the behavior of the $G_{ 1 N}$ element is dominated by $\xi (N)$ leading to $N$ transmission resonances in the conductance of unit height.  Nevertheless, if the effective coupling between the molecule and the lead is much larger than $\gamma^{\rm wire}$, $\xi (N-2)$ will become the dominant term.  As a consequence the conductance spectrum is effectively that of an $(N-2)$--atomic wire~\cite{FCR01a}.  
The resonant behavior inside the wire band ($\labs{\cal E}\rabs \le 1$), and
its modification due to the lead coupling 
is easily understood by writing the
transmission in the following compact exact form valid for all $N \ge 1$
\beann
T=\frac{
4 \delta^2 \sin^2 (\vt)}{
\lrb 
\sin(N+1)\vt 
-\lrb {\delta^2 - \lambda^2}\rrb
\sin(N-1)\vt 
-{2\lambda}
\sin N\vt 
\rrb^2
+4\lrb 
\delta 
\sin N\vt 
-{\lambda \delta}
\sin(N-1)\vt 
\rrb^2
},
\eeann
where the $\sigma= \lambda - \ii \delta = \Sigma / \gamma^{\rm wire}$ is the self energy of the leads (for simplicity assumed equal) normalized by the wire hopping. The parameter $\vt$, defined by
\bea
\label{eq:parametrization}
\vt = \cos^{-1} {\cal E} = \frac{\ii}{2} \ln \frac {{\cal E }- \sqrt{{\cal E}^2 -1}} {{\cal E }+ \sqrt{{\cal E}^2 -1}},
\eea
is real in the wire band giving rise to resonances for injected electrons
matching the wire eigenenergies. Outside the wire band $\vt$ is pure imaginary
($\sin$'s are effectively sinh's) and the transmission as a power law
dependence from energy and an exponential one from wire length, that is
$T \sim \labs 2 {\cal E} \rabs^{-2 N} {\rm ~for~} \labs {\cal E} \rabs \gg1$
in agreement with previous results~\cite{McConnell61}.
This analytic expression for the transmission provide the generalization of
existing ones~\cite{MKR94b,HRHS00,Sumetskii91,MI80} to nonvanishing real part of the self energies.
The density of states ${\cal N} = - {\rm Im~Tr}\lcb G \rcb / \pi $ can also be written in a closed analytical form. One can in fact take advantage of the fact that, due to the wire homogeneity, all the diagonal elements but the first and the last coincide
\beann
\left. \phantom{\frac {\frac 1 2 } {\frac 1 2 }} G_{kk} \; \rabs_{1<k<N}=\frac 1 {\gamma^{\rm wire}} \; \frac
{\xi (N-1) -2 \sigma \xi (N-2) + \sigma^2 \xi(N-3)}
{\xi (N) -2 \sigma \xi (N-1) + \sigma^2 \xi(N-2)} \; .
\eeann
By using the parametrization~(\ref{eq:parametrization}) one can easily
recast the DOS in the following compact form
\beann
{\cal N} = - \frac1 {\pi \gamma^{\rm wire}}
{\rm Im} \; \frac
{N \sin N \vt -2 (N-1) \sigma \sin (N-1)\vt + (N-2) \sigma^2 \sin(N-2)\vt}
{\sin (N+1)\vt -2 \sigma \sin N \vt + \sigma^2 \sin(N-1)\vt} .
\eeann
 \begin{figure}[t]
 \centerline{\epsfig{file=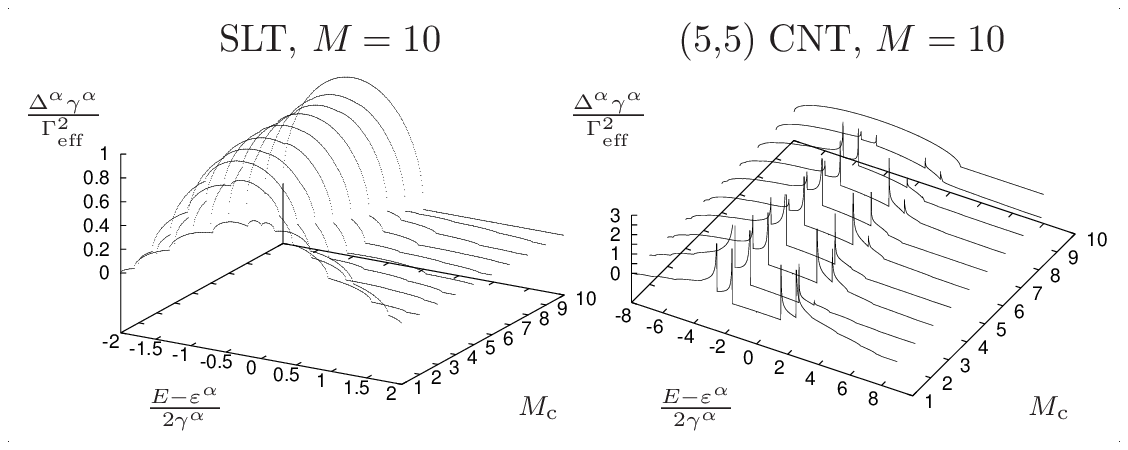, width=0.99\linewidth}}
\phantom{.}\vspace*{1cm}
 \caption{\label{fig:spectrdens} The normalized spectral density as a function of energy and active contacts is plotted for $M=10$ possible atomic contacts available; on--site energies and hopping terms refer to $\alpha=\rm L,R$--leads.}
 \end{figure}
The calculation of the spectral function $\Sigma$, or equivalently 
of the ``surface'' unperturbed lead Green's function is finally needed in getting the conductance.
Due to the form of the interfacial coupling in our model
the self--energy simplifies to
\beann
\Sigma = 
\frac{\Gamma^2_{\rm eff}}{M_{\rm c}}
\sum_{m, m^\prime \le M_{\rm c}} 
 {G}_{\rm lead} \lrb m,m^\prime \rrb ,
\eeann
where only surface terms enter in the sum over the states in the leads, and
the effective coupling is defined as $\Gamma_{\rm eff} \equiv \Gamma \sqrt{M_{\rm c}}$.

In Fig.~\ref{fig:spectrdens}, the spectral function $\Delta = - {\rm Im} \Sigma$ is plotted in the whole range of possible contacts between the SC and MC configuration. 
It is the result of the calculation of the surface Green function obtained
analytically for SLT~\cite{CFR01a,CFR01b} and numerically for CNT~\cite{FCR01a,FCR01b} leads. 
As a function of the number of contacts $M_{\rm c}$, the system interpolates
between two different scenarios. In the MC case, it is effectively
the spectral density of one--dimensional leads, obtained by Newns in
his theory of chemisorption~\cite{Newns69}. 
Only the channel without modulation in the transverse profile of its
wavefunction contributes to transport~\cite{CFR01a,CFR01b}. The two--dimensional character of the leads enters as an energy shift of $2 \gamma^{L,R}$ for SLT and of $\gamma^{L,R}$ for CNT, yielding an asymmetric density profile with respect to the atom on--site energy $\varepsilon^{L,R}$.
In contrast, the SC spectrum is symmetric and richer due to the
contribution of all available channels. 
Additional features characterize CNT leads~\cite{FCR01a,FCR01b}.

In conclusion, we have shown that nanotube / molecular wire / nanotube
hybrid systems exhibit novel features in the conductance. 
The conductance of a homogeneous molecular wire possesses an analytical form
in the full regime of the wire parameters and allows for the insertions of
a nonvanishing real self energy 
typical when considering nanotube leads. 
By tailoring the geometry and dimensionality of the contacts, it is possible to perform a channel selection.
In the MC limit the conductance becomes independent of the topology of the tubular electrodes and transport is effectively one--dimensional.

\end{document}